\documentclass[prl,showpacs,twocolumn]{revtex4}

\usepackage[dvips]{epsfig}
\usepackage{float}

\usepackage{amsmath}
\usepackage{amsfonts}

\newcommand{\begineq}[1]{\begin{equation}\label{#1}}
\newcommand{\eqend}{\end{equation}}

\begin{document}

\title{Anomalous strength of membranes with elastic ridges}

\author{B. A. DiDonna}
\author{T. A. Witten}
\affiliation{James Franck Institute, University of Chicago, Chicago, IL 
60637}


\begin{abstract}{
We report on a 
simulational study of the compression and buckling of elastic ridges formed 
by joining the boundary of a flat sheet to itself.  Such ridges store 
energy anomalously: their resting energy scales as the linear size of the 
sheet to the $1/3$ power.  
We find that the energy required to buckle such a 
ridge is a fixed multiple of the resting energy.  
Thus thin sheets with elastic 
ridges such as crumpled sheets are 
qualitatively stronger than smoothly 
bent sheets. 
}\end{abstract}


\maketitle

\section{Introduction and Theory}
\label{sec:intro}

When the boundaries of an elastic sheet are sufficiently distorted, 
singular deformations in the sheet often occur~\cite{Pomeau,
maha,boudaoud,
Alex,our.stuff,science.paper}.
These singularities 
are immediately apparent when one crumples a sheet of paper between 
the hands. Analogous singularities occur whenever the sheet is forced 
to have points of high curvature~\cite{Alex,our.stuff,science.paper}. 
One can readily form such a 
``puckered'' sheet by fastening the edges of that sheet to form 
disclinations as shown in the cube of Fig.~\ref{fig:grid}. 
Recently it has 
been recognized that such structures typically have their energy 
concentrated into so-called stretching ridges, in which strain and 
curvature energies are in balance~\cite{Alex,our.stuff,science.paper}. 
The ridge increases the stored 
elastic energy in the structure dramatically relative to that of a 
uniformly curved sheet occupying the same region. This enhancement of 
energy grows as a power of the linear size of the structures relative 
to the thickness of the sheet.

The large energy content of these structures suggests that they 
should be strong. They should resist deformation and buckling under 
external loads qualitatively better than nonsingular structures. In 
this note we demonstrate this strength by
showing that the scaling laws 
governing the ridge energy also predict its response to loads and its 
buckling threshold.

We simulate the compression of an elastic cubic shell 
by the application of external, inward-pointing forces at its 
vertices. The cube is formed from a flat sheet as
shown in Fig.~\ref{fig:grid}. With this connectivity
the sheet naturally forms stretching ridges along the edges of 
the cube.
We push on the vertices of the cube until the ridges
buckle into several smaller ridges. 
By rescaling the compressed configuration with the 
same thickness scaling
laws as those appropriate for the resting configuration, we
demonstrate that the response of the ridge to this kind of forcing is
completely described by the
$1/3$ power ridge scaling solution determined in~\cite{Alex}.
We also demonstrate that
the onset of the buckling instability follows this scaling as well.

Ridge scaling comes about as a way to balance bending and stretching 
energy costs. It has been shown~\cite{Alex} that the important 
dimesionless parameter governing the elastic energy of a single ridge 
of length $X$ and material thickness $h$ is the aspect ratio
\begin{equation}
\lambda = \frac{\sqrt{\kappa/Y}}{X} = 
\frac{1}{\sqrt{12 \left( 1 - \nu^2 \right)}} \left(
\frac{h}{X} \right),
\end{equation}
where $Y$ is the $2$-dimensional
Young's modulus of the material, $\kappa$ is an
effective bending modulus, and $\nu$ is its Poisson ratio. 
If all lengths are expressed in units of $X$ and all energies in 
units of $\kappa$, the equilibrium configurations of a thin sheet 
are completely described by a dimensionless form of the 
von K\'{a}rm\'{a}n equations,
\begin{gather}
\nabla^4 f = [\chi, f] + P, \
\lambda^{2}  \nabla^4 \chi = -\frac{1}{2} [f, f],
\label{vk2}
\end{gather}
where all derivatives are taken with respect to
the dimensionless variables
$x/X$ and $y/X$. Here the bracket product represents
\begin{equation}
\left[ a,b \right] = \epsilon_{\alpha\mu} \epsilon_{\beta\nu}
\left( \partial_\alpha \partial_\beta a \right)
\left( \partial_\mu \partial_\nu b \right),
\end{equation}
$P$ is an arbitrary external pressure field acting normal to
the surface (in units of $\kappa X^3$), and 
the potential fields are related to the curvature $C_{\alpha\beta}$
and stress $\sigma_{\alpha\beta}$ by
\begin{gather}
C_{\alpha\beta} = X^{-1} \partial_\alpha \partial_\beta f \\
\sigma_{\alpha\beta} = \kappa X^{-2}
\epsilon_{\alpha\mu} \epsilon_{\beta\nu}
\partial_\mu \partial_\nu \chi.
\end{gather}
This form of $\sigma_{\alpha\beta}$ automatically satisfies
the equilibrium condition
$ \partial_\alpha \sigma_{\alpha\beta} = 0$.

Since $\lambda$ comes into the von K\'{a}rm\'{a}n equations
multiplying the stess source term,
the possible configurations of a thin elastic sheet are well
described by a stress free, $\lambda = 0$ folding solution 
plus boundary layers at the fold lines. 
Lobkovsky's insight in~\cite{Alex}
was to try a scaling solution for the boundary layer
of a single ridge 
which matched the $f$ scaling of the outer, sharp fold solution. For a
fold of dihedral angle $\alpha$ across the line $y=0$, 
$f = \alpha \left| y/X \right|$. Accordingly, on the boundary layer
$f$ should scale with the same power of $\lambda$
as the dimensionless transverse coordinate $y/X$. 
Substituting a scaling form into the dimensionless von
K\'{a}rm\'{a}n equations with $P=0$ 
and equating the highest order terms yields scaling of the form
\begin{equation}
f = \lambda^{1/3} \tilde{f}, \
\chi = \lambda^{-2/3} \tilde{\chi}, \
y = \lambda^{1/3} X \tilde{y}, \
x = X \tilde{x},
\end{equation}
where the tildes denote dimensionless, scale free coordinates and
functions.
This translates to 
$\lambda^{1/3}$ scaling of the boundary
layer width, $\lambda^{-1/3}$ scaling of the transverse ridge curvature, 
and $\lambda^{2/3}$ scaling of the strain along the ridge length.

External forcing applied to the sheet enters the von K\'{a}rm\'{a}n
equations via the term $P$ and via boundary conditions at the
sheet's edges. In this research we consider an external potential
which essentially applies point forces to either end of a ridge. Since
the spatial extent of the applied force is a delta function
we do not expect it to destroy the spatial scaling of the ridge
solution.
Therefore we may reasonably expect to find that the
equilibrium configuration of a ridge under a given compressive force
is identical to rescaled configurations
of ridges with different
material thickness and properly rescaled external force magnitudes. 


To calculate the proper rescaling 
of the forces on the vertices for a similarity
solution, 
we consider our forcing as a boundary condition consisting mainly
of an in-plane point force.
This force amounts to a
point stress at the edge of the sheet with the form
\begin{equation}
\sigma^{(o)}_{xx} = F_o \delta (y).
\end{equation}
So, to find similar scaled configurations of the sheet, we must scale 
$\sigma^{(o)}_{xx}$ the same way $\sigma_{xx}$ scales on the ridge. Since
$\gamma_{xx}$ scales as $\lambda^{2/3}$, and 
$$
\sigma_{xx} \sim Y \gamma_{xx} = \kappa (\lambda X)^{-2} \gamma_{xx},
$$
then $\sigma_{xx} \sim \kappa \lambda^{-4/3} X^{-2} 
\tilde{\gamma}_{xx}$, where $\tilde{\gamma}_{xx}$ is a dimensionless,
scale free function.
To express $\sigma^{(o)}_{xx}$ in similar fashion, we 
substitute the
scale free $y$-variable $\tilde{y} = \lambda^{-1/3} y/X$, so that 
$\delta (y) = \lambda^{-1/3} X^{-1} \delta (\tilde{y})$. Thus, the 
proper scale free force can be written in terms of $F_o$ as
\begin{equation}
F_o = \frac{\kappa \lambda^{-1}}{X} \tilde{F}_o.
\label{eq:ftildeo}
\end{equation}

For reasons which will become clear in the next section, we cannot
measure the force we apply to our ridge with very good
accuracy. However, we can measure
the inward displacement $\Delta$
of the ridge ends caused by this forcing. 
These two quantities may be 
related by assuming that the
work done by equivalent rescaled forces, 
given approximately by $F_o \Delta$,
scales the same as the total energy of the resting ridge
configuration. 
The total energy of a resting ridge scales 
as $\kappa \lambda^{-1/3}$, so equivalent values of $F_o \Delta/ \kappa$
will scale as $\lambda^{-1/3}$. Given the scaling of $\tilde{F}_o$
from Eq.\ref{eq:ftildeo}, the scale free
$\tilde{\Delta}$ must be related to the actual displacement by
\begin{equation}
\Delta = \lambda^{2/3} X \tilde{\Delta}.
\end{equation}


\section{Numerics}
\label{sec:numerics}

\begin{figure}[tbp!]

\center
\epsfig{file=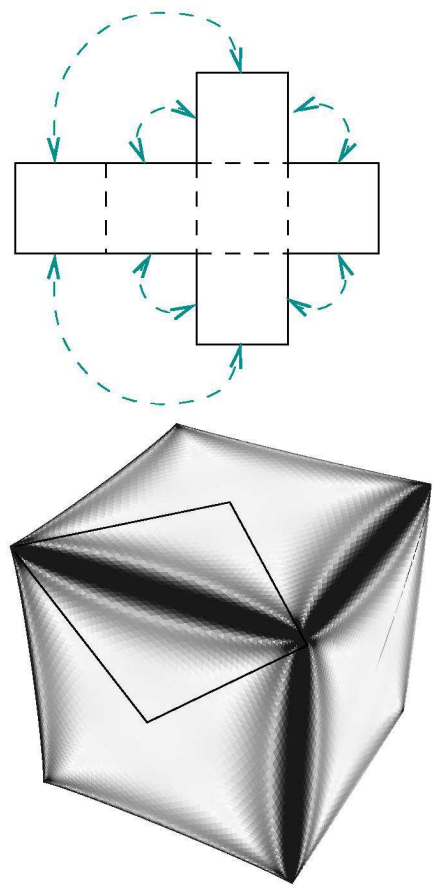} 

\caption{Typical elastic sheet used in this study.  The top image 
illustrates how a flat sheet is joined at its edges to form a cube.  
The bottom 
image shows the resting configuration of the cube 
with no external forces 
acting when the thickness of the sheet is $.0004$ of 
the the edge length and 
the Poisson ratio is $1/3$.  Darker shading represents 
higher strain energy density.  The energy 
was minimized in the outlined diamond-shaped region; the rest of the cube 
shape was inferred by symmetry.  Slight numerical symmetry breaking between 
the left and right sides  of the diamond created slight 
mismatches of the inferred surfaces on other faces, such as the right-hand 
face.  The numerical grid is visible as a quiltlike texture.  It has a finer 
scale at the edges and corners where curvature is larger.}
\label{fig:grid}
\end{figure}


\begin{figure}[tbp!]

{
\scriptsize

\epsfig{file=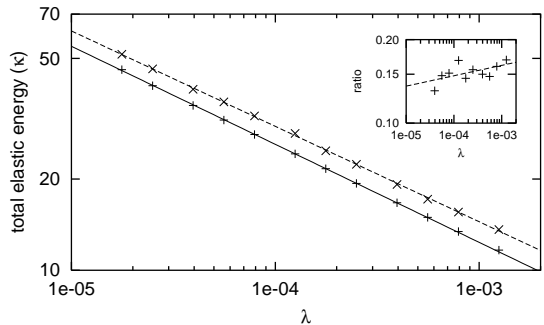}

}
\caption{Energy of ridges at rest and at the buckling threshold. 
Straight lines are least squares fits to a scaling form 
$y=a x^b$.  In this plot 
$\lambda$ ranges from $1.25 \times 10^{-3}$ to $1.77 \times
10^{-5}$. The plot shows the total elastic energy
($E_B+E_S$) in the sheet after minimization.  
The scaling exponent fit for the resting
ridge values (lower line) was $-0.32$, 
the fit at the buckling threshold was $-0.31$. 
The inset shows the
difference between threshold energy and resting energy 
in units of the resting energy. 
This energy ratio is best fit by a scaling
exponent of $0.05 \pm 0.02$
and is consistent with a constant ratio.}
\label{fig:scaling}
\end{figure}

To test these scaling predictions we 
simulate a flat elastic sheet joined to 
form a cube as shown in Fig.~\ref{fig:grid}.  
The sheet is a triangular grid with 
variable grid spacing.  Strains and 
curvature are
taken to be constant across the
face of each triangle, with curvature calculated relative to the 
local normal at each triangle.
Bending and stretching energies were assigned to
the curvature and strains
on each triangle using the forms for elastic energy
presented in~\cite{Seung.Nelson} for a sheet of elastic thickness $h$ 
and Poisson ratio $1/3$.
The gridding was chosen to have smooth gradients in triangle density
over most of the surface while concentrating the
lattice spacing at the vertices by a factor of $10^3$ and across the
ridge-line by a factor of $10^2$ compared to the flat regions far from
the ridge.
The concentrations factors were chosen arbitrarily, 
within the limits of the
mapping, to make the gridding near the vertices as fine as possible.

Pushing on the tips of the cube is accomplished by introducing
repulsive potentials of the 
form $V(r) = C_p/ \left| \vec{r}-\vec{x}_p \right|^2$ 
centered around 
points $\vec{x}_p$ which lie just outside the vertices of the cube. 
To prevent rotation, the vertices are explicitly constrained to
lie on radial lines that pass through the center of the cube 
and the points $(\pm 1,\pm 1,\pm 1)$ which define the vertices of 
a perfect cube.
The center points of the pushing potentials
are located on these radial lines, 
at the point where the vertex would lie if the sheet
were sharply folded at the ridges -- relaxation of the ridge curvature
draws the vertices inward from these points for an unforced resting
ridge. This potential concentrates the forces at the corners without 
creating lattice-scale numerical instabilities.
$C_p$ was varied to apply different loading. 

An inverse
gradient routine~\cite{conj_grad} was used to minimize the total
elastic and potential energy of the sheet as a function of the
coordinates of all the lattice points for given parameters 
$\kappa$,$Y$ and $C_p$.

To save computational time, energy was computed and
minimized on only one ridge-line (the diamond shaped region 
highlighted 
in Fig.~\ref{fig:grid}). The positions of points on the rest 
of the cube were calculated by reflection across symmetry planes. 
This constraint explicitly required that the edges of the 
simulated region be confined to the symmetry reflection planes. 
Thus many forms of 
deformation that break the symmetry of the resting state are not possible in 
the simulation.  However, the minimization may break the left-right symmetry 
of the diamond-shaped region, and we found that it does so to a slight 
degree. The assymmetry shows up in our reconstruction as a slight mismatch 
in other faces of the cube, 
as noted in Fig.~\ref{fig:grid}.  We do not believe these 
mismatches are important for the scaling phenomena we report.


We found minimum energy configurations for ridges of aspect
ratio $\lambda$ ranging from $1.25 \times 10^{-3}$ to $1.77 \times
10^{-5}$. The upper bound on $\lambda$ was determined by the range of
validity of the ridge scaling solution -- 
above this value the width of the ridge becomes
comparable to that the sheet. At the other extreme,
for $\lambda < 10^{-5}$ the radius of
curvature at the ridge line becomes comparable to the spacing of our
lattice and the simulation ceases to be accurate. For each value of
$\lambda$ we first found the minimum energy configuration with no applied
forces. This configuration was then used as the initial condition to find
minimum total energy configurations for the ridge in the presence of 
the tip-pushing potential
described in the numerics section. 
We made the tip-forcing progressively stronger
by increasing the potential coefficient $C_p$ in constant
steps, each time evolving the grid to an energy minimizing
configuration. 
The forcing was increased until the ridge
buckled. Buckling of the ridge was marked by the appearance of new
points of sharp curvature along the ridge as well as 
a sudden decrease in the
total {\it elastic} energy $E_S + E_B$ of the sheet.
Thus the unbuckled sheet is {\it metastable}.
We made no attempt to scale the size of the potential step with the
thickness of the sheet, so while it took nearly $30$ steps to buckle the
ridge in thicker sheets, it only took $6$ or $8$ steps for the thinnest
sheets.

\section{Findings}
\label{sec:findings}

The plot in Fig.~\ref{fig:scaling} shows
scaling of the total elastic energy
in the cube versus $\lambda$ for ridges at rest ($C_{p}=0$) and at 
the buckling threshold.
Scaling of the total elastic energy
for the resting configuration is
consistent with a $\kappa \lambda^{-1/3}$ dependence, in
agreement with prior theory and simulation~\cite{science.paper,Alex}.
Fig.~\ref{fig:scaling} also shows that 
the elastic energy measured 
at the buckling threshhold exhibits {\it exactly the same}
scaling as on resting ridges. This suggests that the response of the
cube to our tip-forcing is completely determined by the ridge,
and that the particular form of our forcing potential does not destroy
the length scaling of the ridge. 
The inset in Fig.~\ref{fig:scaling} shows that that the energy
correction at the buckling threshold is nearly a constant fraction of
the total ridge elastic energy.

\begin{figure}[tbp!]

{

\scriptsize

\epsfig{file=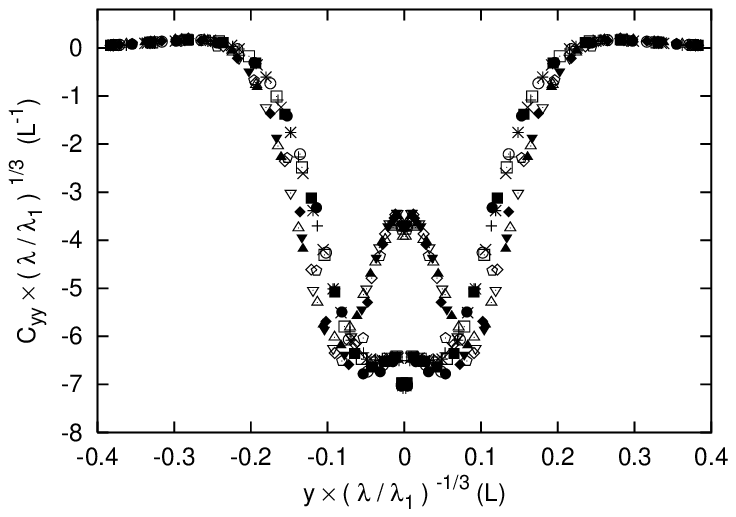}

\vspace{0.2in}

\epsfig{file=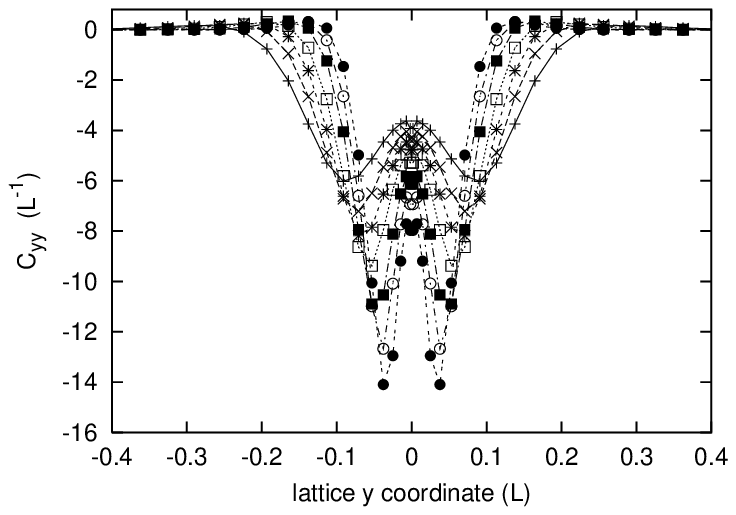}

}
\caption{Demonstration of a similarity solution for the ridge response
to forcing. Both plots show $C_{yy}$, the curvature across the
ridge-line, versus the $y$ material coordinate on the line which bisects
the ridge-line. The data is for sheets with
seven different values of $\lambda$, ranging from 
$1.25 \times 10^{-3}$ to $1.25 \times 10^{-4}$. 
Plot (a) shows
$C_{yy} \times \left( \lambda / \lambda_1 \right)^{1/3}$ vs. 
$y \times \left( \lambda / \lambda_1 \right)^{-1/3}$ for 
the ridges at rest and for ridges with 
inward vertex displacement $\Delta(\lambda) = 
\Delta_1^{(c)} \times \left( \lambda / \lambda_1 \right)^{0.67}$,
where $\Delta$ is measured from the resting vertex
positions, $\lambda_1$ is the aspect ratio for the thickest sheet,
and  $\Delta_1^{(c)}$ was vertex displacement
at the buckling threshold for the
thickest sheet. The profiles with the large central peak are the
buckling threshold values.
(The small dimple 
in the data at $y=0$ is a numerical
artifact due to a discontinuity in the gridding density across the
ridge-line. For finer gridding this dimple goes away, while all other
local values of curvature remain constant.)
Plot (b) shows
unscaled  $C_{yy}$ versus $y$ for 
the buckling threshold profiles plotted in (a).}
\label{fig:rescaling}
\end{figure}

Scaling of the force response is
verified by the demonstration presented in Fig.~\ref{fig:rescaling}
of a similarity solution for ridge shape.
Plot (a) in this figure shows values of $C_{yy}$, 
along a line in the material coordinates which bisects the simulated
ridge line,
for several different sheet thicknesses and two different values
of rescaled ridge tip displacement $\Delta$. 
We tested values
of $\Delta$ which spanned the ridge response from 
zero forcing to the point
of buckling. 
The magnitude of $C_{yy}$ was rescaled by 
$\left( \lambda / \lambda_1 \right)^{1/3}$ and the
$y$ coordinate was rescaled by 
$\left( \lambda / \lambda_1 \right)^{-1/3}$.
We found that the rescaled
$C_{yy}$ profiles mapped onto each other best if 
the values of $\Delta$
used for each sheet thickness were determine by
rescaling $\Delta_1$, the value
used for the thickest sheet, by 
$\left( \lambda / \lambda_1 \right)^{0.67}$, where $\lambda_1$ is the
aspect ratio of the thickest sheet.  
For comparison, the
unscaled $C_{yy}$ versus $y$ for a particular rescaled $\Delta$ is
shown in Fig.~\ref{fig:rescaling}(b).
The rescaling exponents used for $C_{yy}$ and $y$ were taken from the
ridge scaling solution.
The $\Delta$ rescaling exponent is very close to the theoretical
value of $2/3$ derived above. 


We have demonstrated scaling of the ridge response to a typical form of
external forcing. Since the buckling threshold of the ridge under
external forcing is determined by the energy scaling of the resting
ridge solution, the relatively high elastic energy content of very thin
ridges translates directly into increased ridge strength. 
We are actively pursuing a better understanding of the buckling properties 
of this distinctive elastic structure.

\acknowledgments{
This work was supported in part by the National Science Foundation under 
Award number DMR-9975533.}


\end{document}